# Experimental Study of the Inductance of Pinned Vortices in Superconducting $YBa_2Cu_3O_{7-\delta}$ Films

Aaron A. Pesetski and Thomas R. Lemberger

*Department of Physics*

*The Ohio State University*

*Columbus, OH 43210-1106*

Using a two-coil mutual inductance method, we have measured the complex resistivity, $\rho_V(T,B_e)$, of pinned vortices in c-axis pulsed laser deposited $YBa_2Cu_3O_{7-\delta}$ films with magnetic fields applied perpendicular to the film. At low frequencies, (<100 kHz), $\rho_V$ is inductive and is inversely proportional to the Labusch parameter, the average vortex pinning force constant, $\kappa_{exp}$. The observed weakening of $\kappa_{exp}$ with $B_e$ is consistent with a simple model based on linear pinning defects. Adding classical thermal fluctuations to the model in a simple way describes the observed linear T dependence of $\rho_V$, below ~15 K and provides reasonable values for the effective radius ($\approx$ 3 Å to 8 Å) of the defects and the depth of the pinning potential. The success of this model implies that thermal supercurrent (phase) fluctuations have their full classical amplitude down to 5 K for frequencies below the characteristic depinning frequency. To date, no sufficient theory exists to explain the data between ~15 K and the vortex glass melting temperature.

I. Introduction

This paper reports measurements and a comprehensive analysis of the complex resistivity of vortices in the vortex glass phase in $YBa_2Cu_3O_{7-\delta}$ films. The present work provides a more complete study of the dependence of the resistivity on temperature, T, and applied field, $B_e$, and a more detailed analysis, than previous studies.[1,2] The measurements were performed in a uniform external magnetic field of up to 6 Tesla at temperatures between 5 K and 100 K. The motivation is to develop a sufficiently accurate understanding of vortex pinning that vortex behavior can be used to probe the intrinsic properties of the material. In this regard, this study is analogous to recent measurements of the transverse thermal conductivity in a magnetic field.[3] This study relies on the work of Clem and Coffey[4] for interpretation of the two-coil mutual inductance measurements in terms of vortex parameters and the theory of Nelson and Vinokur[5] for the behavior of vortices pinned to extended defects.

In the ranges of temperature and field of interest here, vortices are pinned and their complex resistivity is predominantly inductive. Pinning results from as yet unidentified lattice defects that form naturally during film growth. The vortex resistivity is deduced from the mutual inductance of coils located on opposite sides of the film. An AC current in the primary coil induces a small supercurrent in the film which oscillates the vortices about their equilibrium positions. Through the dependence of the vortex inductance on temperature and vortex density, the experiment probes the first stages of depinning of vortices by thermal fluctuations and intervortex repulsion.

The inductance of pinned vortices has received much less experimental[1-2,6-13] and theoretical[4,14] attention than nonlinear current-voltage characteristics and flux creep phenomena. As the present work shows, it is possible to obtain important information from the vortex inductance that cannot be obtained from these other measurements. The important distinction is that in a measurement of the vortex inductance, the perturbing AC supercurrent displaces an average vortex only a few hundredths of an Angstrom from its equilibrium configuration, whereas measurements of nonlinear current voltage characteristics and flux creep involve hopping of vortices from pinning site to pinning site. The former probes the shape of the pinning potential near its bottom, while the latter probes the height and width of the energy barrier between nearby pinning sites.

Early on, Hylton and Beasley[15] recognized that the critical current density of about $10^7$ A/cm$^2$ measured in YBCO films implies that vortices are strongly pinned. To provide some insight into the nature of the lattice defect necessary to provide such strong pinning, they assumed the existence of point pinning centers which extinguish superconductivity within a radius equal to the ab-plane coherence length, $\xi_{ab}$. Even with such strong pinning centers, Hylton and Beasley deduced that a full quarter of the length of each vortex would have to be pinned. They estimated a mean spacing between defects of about 34 Å, which implies that a large fraction of the film is, in fact, not superconducting. They noted that this provides a possible explanation for the lower superfluid densities observed in films when compared to single crystals.

Based on our data and the following considerations, we conclude that as regards vortex inductance, pinning sites are more like continuous, one dimensional extended defects than point defects. First, if three-fourths of each vortex were unpinned, motion of unpinned segments in response to the AC probe field would provide a vortex inductance larger than is observed, regardless of the strength of the point pins. Second, within the Hylton and Beasley model itself, if the point pins are assumed to be somewhat weaker, e.g., their effective radius is $\xi_{ab}/2$, then vortices would have to intersect one of these defects in nearly every unit cell layer of the film. This would require either a very large density of point defects, or, more likely, point defects that are highly correlated along the c-axis, i.e., a string of correlated point defects which would constitute a continuous one-dimensional pinning site similar to the columnar defects generated by ion bombardment, but not necessarily rectilinear. Finally, Diaz *et al* observed a strong angular dependence to the critical current in YBCO films indicating that it is much easier to depin vortices when the applied field is tilted by a few degrees.[16] Again, this implies that extended defects are responsible for pinning.

Further evidence favoring extended defect pinning was provided by Golosovsky, who, in his review,[2] noted that pinning in YBCO films is much stronger than can be accounted for by the collective pinning model, which is based on point defects, and is constructed to describe reasonably clean single crystals.[17] Nevertheless, Golosovsky noted that the T-linear increase in vortex inductance observed in his experiments mirrored the T-linear decrease in the depth of the effective pinning potential that is calculated within the collective pinning model. The present work confirms T-linear behavior and argues that it results from thermal fluctuations, as Golosovsky speculated, but it is more accurately described by the Nelson-Vinokur[5] model for pinning by extended defects than by the collective pinning theory.

Regarding the dependence of the vortex inductance on vortex density, Xenikos et al.[1] noted that the field dependence of the vortex inductance in YBCO films is much weaker than predicted by the collective pinning model. They speculated that vortex pinning might be better modeled with extended pinning sites. Our data support this conclusion. The observed field dependence can be explained by introducing vortex interactions into an extended defect model.

In summary, the objective of the present work is to examine in quantitative detail the inductance of pinned vortices as a function of T and $B_e$. The data are examined in the context of an idealized classical thermal model in which pinning arises from linear extended defects in the crystal lattice, and is weakened by thermal fluctuations and vortex interactions.

## II. Experimental Setup

The mutual inductance apparatus used to measure the vortex inductance is described in detail in Refs. 18 and 19. An unpatterned YBCO film is placed between two coaxial, quadrupole coils. The magnitude and phase of the mutual inductance between the coils are recorded as functions of T and $B_e$. All data are taken in the linear response regime, that is, the mutual inductance is independent of the amplitude of the current in the primary coil. Data typically are taken at 50 kHz to obtain good signal-to-noise, but the results are independent of frequency down to 500 Hz. The coils and film sit in the bore of a 6 Tesla superconducting solenoid which is coaxial with the coils. The data presented are zero-field cooled. However, no difference is observed in field cooled measurements for $B_e > 1$ T. The complex resistivity is extracted from the complex mutual inductance data through a numerical model of the experimental setup.[4,20] A detailed analysis of the experiment has been performed to ensure that this method is accurate.[20,21] Vortex pinning parameters are extracted using the Clem and Coffey[4] model as discussed below.

Data are presented on two 500 Å thick $YBa_2Cu_3O_{7-\delta}$ films grown on 1 cm$^2$ LaAlO$_3$ substrates by pulsed laser deposition (PLD). The films have their c-axes aligned perpendicular to the substrate and parallel to the field. They are typical PLD YBCO films: $T_C$ = 88.9 K and $\lambda_{ab}$(T=0,B=0) = 1970Å for film 1, and $T_C$ = 84.4 K and $\lambda_{ab}$(0,0) = 2500Å for film 2. Both films have critical current densities of about 10$^7$ A/cm$^2$ at 5 K, also typical of PLD films. The vortex glass melting curves are consistent with $B_g \propto (1 - T_g/T_C)^{4/3}$, where $T_g$ is the vortex glass melting temperature and $B_g$ is the vortex glass melting field, which is consistent with previous measurements[17,18,22]. Preliminary measurements on films made using other methods, i.e., sputtering and coevaporation with post annealing, behave somewhat differently, presumably due to differences in pinning sites.

### III. Results

Before the introduction of vortices ($B_e$ = 0), and for T < $T_C$ – 1 K, the film's complex resistivity, $\rho$(T,$B_e$=0), is dominated by the inductivity of the superfluid, i.e., $\rho(T,0) = i\mu_0\omega\lambda_{ab}^2(T)$, where $\lambda_{ab}$ is the magnetic penetration depth and $1/\lambda_{ab}^2(T)$ is proportional to the superfluid density, $n_s(T)$. Dissipation resulting from single-particle excitations is undetectable. When vortices are present and T is less than $T_g$, $\rho(T,B_e)$ is again predominantly inductive at our measurement frequencies. The film resistivity is conveniently expressed in terms of an effective penetration depth, $\lambda_{eff}^2(T,B_e) \equiv \rho_2/\mu_0\omega$. Figures 1a and 1b show $1/\lambda_{eff}^2$ vs. T for 5 K ≤ T ≤ 100K and 0 ≤ $B_e$ ≤ 6 Tesla. The dashed lines for T < 5 K are extrapolations based on polynomial fits to the data between 5K and 30K and are used to extract $1/\lambda_{eff}^2$(T,$B_e$=0).

Examination of $1/\lambda_{eff}^2(T,B_e=0)$ reveals that the superfluid density, $n_s(T) \propto 1/\lambda_{eff}^2(T,B_e=0)$, is flat at low T, presumably due to residual disorder in the film rounding off the T-linear behavior of a d-wave superconductor. The other curves show that $\lambda_{eff}^2$ increases when either T or $B_e$ increases. Note that for $B_e \geq 1$ Tesla, $\lambda_{eff}^2$ increases linearly with T at low T, even though $n_s(T)$ is flat. This behavior will lead to the important conclusion that classical thermal fluctuations dominate the T dependence of vortex pinning.

We isolate the inductivity of the vortices by subtracting away that of the superfluid, following the theory of Clem and Coffey (CC).[4] CC derived an integral expression for the mutual inductance of coaxial coils on opposite sides of a superconducting film in the presence of a magnetic field. The mutual inductance is a function of the Labusch parameter (the linear restoring force constant), $\kappa_0$, and the vortex viscosity, $\eta$, which characterize the motion of vortices. CC delineated conditions under which it is sensible to ignore the effects of the nonuniform current distribution created by the drive coil and extract $\kappa_0$ and $\eta$ directly from $\rho(T,B_e)$. These conditions are satisfied in our film.

CC treated an idealized case where, in the absence of the perturbation introduced by the weak, inhomogeneous field from the primary coil, vortices are arranged on a perfect lattice, and all vortices are centered on identical pinning sites. At fields below about 1 Tesla, the experimental situation is complicated by the inhomogeneous entry of vortices into the film along paths of least resistance. These complications are discussed in section VII. For $B_e > 1$ Tesla, the vortex density is nearly uniform and pinning sites can be treated as identical sites with a Labusch parameter, $\kappa_0(T,B_e)$. The experimentally observed Labusch parameter, obtained using the CC model, is thus an average over different pinning strengths and different displacements form equilibrium resulting from lattice distortions, and thermally induced motion.

As stated above, the film's complex resistivity is completely inductive at low frequencies except for temperatures very close to $T_g$, yielding $\rho(T,B) = i\omega\mu_0\lambda_{eff}^2$. The vortex contribution to the film resistivity is characterized by the Campbell penetration depth, $\lambda_C$, defined by $\lambda_C^2 \equiv \lambda_{eff}^2 - \lambda_\omega^2$, where $\lambda_\omega$ is the penetration depth of the superfluid in the presence of a magnetic field. In an s-wave superconductor, the superfluid density is suppressed in and near the cores of the vortices. The average superfluid density is reduced to roughly the volume fraction not occupied by vortices, $1 - B_e/B_{C2}(T)$ yielding $\lambda_\omega^2(T,B_e) = \lambda_{ab}^2(T)/(1 - B_e/B_{C2})$. In a d-wave superconductor this suppression extends well beyond the vortex cores. The supercurrents surrounding the core of the vortex extend to a distance $\lambda_{ab}$. In directions where the supercurrent is parallel to the nodes of the order parameter, the superfluid density is suppressed. The total suppression is $\lambda_\omega^2(T,B_e) = \lambda_{ab}^2(T)/\left(1 - \sqrt{B_e/B_{C2}}\right)$. In our films, $B_e << B_{C2}$ and thus $\lambda_\omega(T) \approx \lambda_{ab}(T)$ except near $T_C$. As a result, the exact functional form of $B_{C2}(T)$ is only relevant near $T_C$ and in analyzing our data, we use the approximation $B_{C2}(T) \approx (2\text{ Tesla/K})(T_C-T)$.

With these definitions, $\lambda_C^2/B_e$ is proportional to the inductance per vortex. Figure 2 shows $\lambda_C^2(T,B_e)/B_e$ *vs*. T for both films at various magnetic fields. The extrapolations to T < 5 K are derived from the extrapolations in Figs.1a and 1b. $\lambda_C^2/B_e$ increases linearly with T at low T, then diverges at $T_g$, the vortex glass melting temperature. Fitting the divergence at $T_g$ shows that $\lambda_C^2$ diverges as $(1 - T/T_g)^{-1.0 \pm 0.1}$.
Above $T_g$, vortices are mobile, and their resistivity has a large dissipative component.
Figure 3 shows that the melting curves agree well with previous work on films,[17,18,22]
with $B_g \propto (T_C - T_g)^{4/3}$ as predicted by vortex glass melting theory.[19]

The experimental Labusch parameter, $\kappa_{exp}$, is derived from $\lambda_C^2/B_e$.

$$\kappa_{exp} \equiv \frac{\Phi_0 B_e}{\mu_0 \lambda_C^2} \qquad (1)$$

Figures 4a and 4b show $\kappa_{exp}$ vs. T. The values of $\kappa_{exp}$ at T = 0 are typical for PLD YBCO films.[1-2,6-13] Note that for all fields $\kappa_{exp}$ decreases linearly in T at low temperature, then vanishes as $T \rightarrow T_g$. Our experiment was conducted at frequencies too low to observe the dissipative part of the vortex resistivity and thereby measure the vortex viscosity, $\eta$. $\eta$ has been measured in a number of microwave experiments on similar films,[2,9-13,22,23] and in the remainder of this paper we use an average value of $\eta(T=0) \approx 1 \times 10^{-6}$ Ns/m$^2$. This value yields a characteristic vortex depinning frequency $\omega_p/2\pi \sim 40$ GHz, well above our ~ 50 kHz operating frequency.

IV. Pinning Model

Our goal is to extract quantitative information about the pinning force experienced by each vortex from the dependence of $\kappa_{exp}$ on T and $B_e$. To do so we need to model how the intrinsic Labusch parameter, $\kappa_0(T)$, is effectively reduced by thermal fluctuations and vortex interactions. To do so, we begin by considering an idealized situation in which the vortices form a perfect lattice and each vortex sits in an identical pinning potential with a Labusch parameter $\kappa_0$, then add the effects of thermal fluctuations and vortex interactions.

Consider the behavior of an isolated vortex pinned in an extended defect at T = 0. Assuming that the pinning potential, U(T,ρ), is azimuthally symmetric with respect to the radial displacement, ρ, we expect U to have the form: $U_0(T)f(\rho/\xi_{ab}(T))$, where $U_0(T)$ is the depth of the pinning potential and $f(\rho/\xi_{ab}(T))$ ranges from 0 at ρ = 0 to unity at ρ ~ $2\xi_{ab}(T)$ with an inflection point near ρ = $\xi_{ab}(T)$. Since the exact shape of the pinning potential is not known we will approximate it with a polynomial in ρ for ρ < $\xi_{ab}(T)$:

$$U(T,\rho) = \frac{1}{2}\kappa_0(T)\xi_{ab}^2(T)\left(\frac{\rho}{\xi_{ab}(T)}\right)^2 - \frac{1}{12}\kappa_0(T)\xi_{ab}^2(T)\left(\frac{\rho}{\xi_{ab}(T)}\right)^4 + O(\rho^6). \quad (2)$$

The coefficient of the second order term is chosen such that the force constant at ρ = 0 is $\kappa_0(T)$. The coefficient of the fourth order term is chosen to such that the inflection point is located at ρ = $\xi_{ab}$.

Eq. (2) is a good approximation to the true pinning potential even for ρ ≈ $\xi_{ab}(T)$. To demonstrate this, note that the maximum force which the potential can generate is $F_{max} = {}^2\!/_3\,\kappa_0\xi_{ab}$. The largest current the film can carry is thus

$$J_C = \frac{2\kappa_0\xi_{ab}}{3\Phi_0} \approx 1\times10^7 \text{ A/cm}^2.$$

We can measure the critical current experimentally by ramping the current in drive coil of our apparatus into the nonlinear response regime, and calculating the induced current density at the onset of nonlinearity. Doing so yielded critical currents of $2\times10^7$ A/cm² and $4\times10^6$ A/cm² for films 1 and 2, respectively, which are typical values for PLD YBCO films and show good agreement with our model.

We can estimate $\kappa_0(T)$ by calculating the pinning energy, the energy saved by locating a vortex on a defect. If we assume that the pinning defect is effectively a cylindrical hole of radius $r_d$, the pinning energy per unit length is approximately $\dfrac{B_C^2(T)}{2\mu_0}\pi r_d^2$, where $B_C(T) = \dfrac{\Phi_0}{2\sqrt{2}\pi\lambda_{ab}(T)\xi_{ab}(T)}$ is the thermodynamic critical field.

Using Eq. (2), the depth of the pinning potential is roughly $2U(\rho=\xi_{ab}) = \tfrac{5}{6}\kappa_0 \xi_{ab}^2$. Equating this to the pinning energy, we get

$$\kappa_0(T) \approx \frac{3\Phi_0^2 r_d^2}{40\pi\mu_0 \lambda_{ab}^2(T)\xi_{ab}^4(T)}. \qquad (3)$$

Eq. (3) can be used to estimate the radius of the pinning defect from the measured Labusch parameter, $\kappa_{exp}(T\to 0, B\to 0) \approx 3\times 10^5$ N/m$^2$. Assuming $\xi_{ab}(T=0) = 15$ Å yields $r_d \approx 8$ Å, which is about half of $\xi_{ab}(T=0)$, a reasonable number in light of Hylton and Beasley's analysis.

In the above analysis, we considered only the condensation energy, $\dfrac{B_C^2(T)}{2\mu_0}$, saved in locating the vortex on the pinning site. In addition to the condensation energy, there is a substantial kinetic energy associated with the supercurrents in and around the vortex core. With such a large defect, $r_d \approx 8$ Å, a portion of the kinetic energy will also be saved by locating the vortex on the defect. If all of the kinetic energy were saved, the pinning energy would be larger by a factor $\ln\left(\lambda_{ab}/\xi_{ab}\right) \approx 5$, and thus the estimated defect radius would be $r_d \approx 3.5$ Å. Since it is unclear what fraction of the kinetic energy is saved the true radius is somewhere between 3 Å and 8 Å.

Eq. (3) shows that this model leads to a temperature dependence for the intrinsic Labusch parameter of $\kappa_0(T) \sim 1/\lambda_{ab}^2(T)\xi_{ab}^4(T)$. If we assume that the pinning defect suppresses superconductivity within a distance proportional to $\xi_{ab}(T)$ through the proximity effect, then $\kappa_0(T) \sim 1/\lambda_{ab}^2(T)\xi_{ab}^2(T)$. The dashed curves in Figs. 4a and 4b are $\kappa_0(T)$ calculated by using the measured values $\kappa_{exp}(T=0,B=0)$ and $\lambda_{eff}(T,B=0)$ and $\xi_{ab}(T)$ calculated by Ulm *et al*[24] for disordered YBCO. To allow for an estimated fluctuation-induced suppression of $T_C$ of about 6 K, $\xi_{ab}(T)$ is calculated such that it diverges at $T/T_C = 1.07$. As is clearly evident in Figs. 4a and 4b, $\kappa_{exp}(T)$ differs dramatically from $\kappa_0(T)$. (The difference is even more severe when $\kappa_0(T) \sim 1/\lambda_{ab}^2(T)\xi_{ab}^2(T)$ is used.) The difference arises primarily from the reduction of $\kappa_0(T)$ resulting from classical thermal fluctuations of the vortex position as discussed below.

## V. Temperature Dependence of the Labusch Parameter

To understand the effects of thermal fluctuations in the simplest way, we continue to consider isolated vortices. In section VII we will show that the net force resulting from vortex interactions is relatively weak in our films and neglecting them is reasonably accurate at low temperatures and at fields near 1 Tesla. At T = 0 and with no applied supercurrents, each vortex sits at the center of its pinning site. For T > 0, the phase of the order parameter within each unit cell thick layer fluctuates rapidly in space and time, independent of the presence of vortices. The gradients in the phase of the order parameter resulting from these fluctuations represent supercurrents which drive the vortex through a random path about the minimum of the pinning potential. Since the thermal fluctuations occur up to frequencies much higher than those of the measurement, the measured pinning force is an average of the curvature of the pinning potential over the path of the wandering vortex. Rather than attempting to calculate this average, we will approximate the effective Labusch parameter with the curvature of the pinning potential, $U(\rho)$, in the direction of the force created by the primary coil, at the *rms* displacement from the center of the pinning potential, $\rho_{rms} = <\rho^2>^{1/2}$. For simplicity we choose our coordinate system such that the x direction is the direction of the force produced in our measurement. This yields,

$$\kappa = \left.\frac{\partial^2 U}{\partial x^2}\right|_{\rho=\rho_{rms}} = \kappa_0 \left[1 - \frac{1}{3}\left(\frac{\rho_{rms}}{\xi_{ab}}\right)^2 \left(1 + 2\cdot\cos^2\phi\right)\right], \qquad (4)$$

where $\phi$ is the angle the displacement makes with respect to the x axis. Each vortex will be displaced in a different direction so we must average over $\phi$. Since it is the inductance of the vortices which is additive, we must average $1/\kappa$, not $\kappa$. From this average we obtain the effective Labusch parameter,

$$\kappa_{eff} = \left[\frac{1}{2\pi}\int_0^{2\pi} \kappa^{-1}(\rho_{rms})d\phi\right]^{-1} = \kappa_0\sqrt{1 - \frac{4}{3}\left(\frac{\rho_{rms}}{\xi_{ab}}\right)^2 + \frac{1}{3}\left(\frac{\rho_{rms}}{\xi_{ab}}\right)^4} \underset{\rho_{rms}\ll\xi_{ab}}{\approx} \kappa_0\left[1 - \frac{2}{3}\left(\frac{\rho_{rms}}{\xi_{ab}}\right)^2\right]. \qquad (5)$$

We can calculate $\rho_{rms}$ by appealing to the equipartition theorem. In order to use the equipartition theorem, we must calculate the total energy increase when the vortex is displaced. Consider a segment of vortex of length $\ell_z$ which moves independently of the segments above and below it. $\ell_z$ depends on the flexibility of the vortex. If each vortex were a rigid rod, then $\ell_z$ would equal the film thickness, d. If the YBCO crystal were completely decoupled, then $\ell_z$ would be the c-axis lattice constant. The average pinning energy of each vortex segment is $\langle U(\rho)\rangle \ell_z$, where $\langle U \rangle$ is the average of U over the wandering path of the vortex. We can approximate the increase in potential energy with $\langle U(\rho)\rangle \ell_z \approx U(\rho_{rms})\ell_z$.

In addition to increasing the potential energy, thermal vortex motion also increases the length and therefore the line energy of the vortex. If the vortex segment at

z = 0 has moved a distance $\rho_{rms}$ from the center of the potential, and the adjacent segment at z = $\ell_z$ has moved a distance $\rho_{rms}$ in another direction, then the added length, averaged over all angles, is roughly:

$$\delta\ell = \frac{1}{2\pi}\int_0^{2\pi}\sqrt{2\rho_{rms}^2(1-\cos\phi)+\ell_z^2}\,d\phi - \ell_z$$

$$= \ell_z\left[\frac{2}{\pi}\sqrt{1+\left(\frac{2\rho_{rms}}{\ell_z}\right)^2}\,E\left(\sqrt{\frac{4\rho_{rms}^2}{\ell_z^2+4\rho_{rms}^2}}\right)-1\right] \underset{\rho_{rms}\ll\ell_z}{\approx} \frac{\rho_{rms}^2}{\ell_z}, \quad (6)$$

where E(x) is the complete elliptic integral of the second kind. The line energy per unit length is $\varepsilon_1 = \frac{\Phi_0^2}{4\pi\mu_0\lambda_{ab}^2}\ln\left(\frac{\lambda_{ab}}{\xi_{ab}}\right)$ for an isotropic material.[25] In the present case, the additional line length is in the ab-plane. Because of the anisotropy in $\lambda$ in YBCO, the line energy in the ab-plane, $\tilde{\varepsilon}_1$, is reduced from $\varepsilon_1$ by a factor $M_{ab}/M_c \approx 1/40$.[5]

According to the equipartition theorem each independently fluctuating vortex segment has an average thermal energy of $k_BT$. Equating this with the total increase in energy yields

$$k_BT = \langle U(\rho)\rangle\ell_z + \tilde{\varepsilon}_1\delta\ell \approx \left[\frac{\kappa_0\rho_{rms}^2}{2} - \frac{\kappa_0\rho_{rms}^4}{12\xi_{ab}^2}\right]\ell_z + \frac{\rho_{rms}^2}{\ell_z}\tilde{\varepsilon}_1. \quad (7)$$

We can calculate $\rho_{rms}$ to lowest order in T by assuming that each of the two terms on the right hand side of Eq. (7) has an average value of $\frac{1}{2}k_BT$:

$$\rho_{rms}^2 = 3\xi_{ab}^2\left[1-\sqrt{1-\frac{2}{3}\frac{k_BT}{\ell_z\kappa_0\xi_{ab}^2}}\right] \underset{T\to 0}{\approx} \frac{k_BT}{\ell_z\kappa_0}. \quad (8)$$

This assumption will be justified later. We can now calculate the average Labusch parameter using Eq. (5):

$$\kappa_{eff}(T) \approx \kappa_0(T)\left[1 - \frac{2}{3}\left(\frac{\rho_{rms}(T)}{\xi_{ab}(T)}\right)^2\right] = \kappa_0(T) - \frac{2k_B T}{3\ell_z \xi_{ab}^2(T)}. \qquad (9)$$

Since $\kappa_0(T)$ and $\xi_{ab}(T)$ are both roughly constant at low temperatures, Eq. (9) predicts a T-linear decrease in $\kappa_{eff}$ at low T, as is observed. Identifying $\kappa_{exp} \approx \kappa_{eff}$, and using the experimental values $\left.\frac{d\kappa_{exp}}{dT}\right|_{T \to 0} \approx -5.5 \times 10^3$ N/m$^2$K for film 1 and $-4.0 \times 10^3$ N/m$^2$K for film 2 (Figs. 4a & 4b), we find $\ell_z(T=0) \approx 8$ Å for film 1 and $\ell_z(T=0) \approx 10$ Å for film 2. These values of $\ell_z(T=0)$ are surprisingly small. They imply that the vortex segment within each YBCO unit cell layer is decoupled from the segments above and below it. To gauge the uncertainty in $\ell_z$, we have repeated the above analysis with several plausible, analytically tractable pinning potentials which have a curvature $\kappa_0$ at $\rho = 0$ and an inflection point at $\rho = \xi_{ab}$, [e.g.

$$\frac{4\kappa_0 \xi_{ab}^2}{\pi^2}\left[1 - \cos\left(\frac{\pi\rho}{2\xi_{ab}}\right)\right], \quad 4\kappa_0 \xi_{ab}^2\left[1 - e^{-\frac{\rho^2}{2\xi_{ab}^2}}\right], \quad \frac{\frac{1}{2}\kappa_0 \rho^2}{1 + \rho^2/3\xi_{ab}^2}, \quad c^2 \kappa_0 \xi_{ab}^2\left[1 - \mathrm{sec}\,h\left(\frac{\rho}{c\xi_{ab}}\right)\right].]$$

Values obtained for $\ell_z$ by repeating the procedure used above with one of these potentials ranged from 9 Å to 20 Å, which are much more appealing values. They are very close to both $\xi_{ab}(T=0)$ and the c-axis lattice constant for YBCO, 11.7 Å. This wide range of values implies that the value of $\ell_z$ is sensitive to the exact functional form of the pinning potential.

An expression for the fluctuation length, $\ell_z$, follows from our assertion that the two terms in Eq. (7) are equal. For $\rho_{rms} \ll \xi_{ab}$, i.e. T $\to$ 0, we can equate the lowest order terms in the pinning potential and the added line energy to obtain:

$$\ell_z = \sqrt{2\frac{\tilde{\varepsilon}_1}{\kappa_0}} = \sqrt{2\frac{M_{ab}}{M_c}\ln\left(\frac{\lambda_{ab}}{\xi_{ab}}\right)\frac{\Phi_0^2}{4\pi\mu_0\kappa_0\lambda_{ab}^2}}. \qquad (10)$$

Using $\kappa_0(T=0)$ from Fig. 4 and $\lambda_{ab}(T=0)$ from Fig. 1 yields $\ell_z(T=0) \approx 25$ Å.

This apparently *ad hoc* procedure of equating the two energy terms in Eq. 7 to calculate $\ell_z$ can be justified by considering the Fourier transform of the vortex position in the ab-plane, $\vec{\rho}(z)$, with respect to z. For small wave vectors, the energy is dominated by the pinning potential and $\vec{\rho}(q_z)$ is constant while for large wave vectors, the energy is dominated by the line energy and the $\vec{\rho}(q_z)$ is very small. The cut off value, $1/\ell_z$, occurs when the pinning energy is equal to the added line energy resulting in $\ell_z$ given by Eq. (10). This method of Fourier transforming the vortex position is similar to that used in the formalism derived by Nelson and Vinokur which is described in greater detail in Ref. 5.

Our model fails to describe the observed Labusch parameter at temperatures above 15 K. Fig. 5 shows three curves: $\kappa_{exp}(T,B_e=1T)$, the experimental Labusch parameter for film 1 taken from Fig. 4a, $\kappa_0(T)$, the intrinsic Labusch parameter of the pinning site calculated in Eq. (3), and $\kappa_{eff}(T)$ the effective Labusch parameter which results when the effects of thermal fluctuations are included. For the sake of comparison, we assume that 1 Tesla is sufficiently small that vortex interactions are negligible and sufficiently large that the vortex density is uniform, and thus $\kappa_{exp}(T)$ is well described by the isolated vortex model. Note that $\kappa_{eff}(T)$ deviates from $\kappa_{exp}(T)$ at about 15 K, the same point where $\kappa_0(T)$ experiences a sharp downward curvature. When the analytic forms for $U(T,\rho)$ discussed previously are used to calculate $\kappa_{eff}(T)$, the same deviation is observed, although at a slightly higher temperature. This suggests that $\kappa_0(T)$ is much flatter than predicted. However, even if $\xi_{ab}(T)$ is assumed to be constant for all T, and $\kappa_0(T)$ had only the temperature dependence of $\lambda_{ab}^{-2}(T)$, it would still be sufficient to generate a discrepancy between $\kappa_{eff}(T)$ and $\kappa_{exp}(T)$. This discrepancy and the upward curvature in $\kappa_{exp}(T)$ above ~20 K are not understood.

Another way of illustrating this problem is by considering the pinning energy of a single fluctuating segment. Using the values given above, the characteristic pinning energy is $U_0(T=0)\ell_z(T=0)/k_B \approx 50$ K. Presumably, as T increases, the pinning energy decreases and we would expect the vortex to become unpinned somewhere below 50 K. (Note that $\kappa_{eff}(T)$ drops to zero at about 44 K.) In order for the vortex to remain pinned until well above 80 K, as observed in the data, the pinning energy of a single segment must increase with temperature. For this to occur, either the pinning potential must increase with temperature, or the fluctuation length, $\ell_z$, must increase rapidly enough to compensate for the decreasing $U(\rho,T)$. Both of these possibilities seem unlikely. Clearly more theoretical work is needed to understand the high temperature behavior.

## VI. Thermal Supercurrent Fluctuations

Thermal motion of vortices is caused by fluctuations in the supercurrent density in the ab-plane of each unit cell of the film. Knowing the thermal motion of the vortices, we can work backwards to deduce some properties of the thermal supercurrents. First, the data indicate that vortex segments on the order of 10 Å long fluctuate independently at 5 K. Thus, given the unit cell thickness of c = 11.7 Å, the supercurrents in each copper oxide layer of the film must be uncorrelated. YBCO should be quasi-two-dimensional so far as these fluctuations are concerned.

To estimate the magnitude of thermal supercurrents in each unit cell layer, we equate the mean square force they exert on a vortex, $<J_s^2>\Phi_0^2$, to the mean square restoring force exerted by the pinning potential, $\kappa_0^2<\rho^2>$ replacing $\ell_z$ with c, the c-axis lattice constant. We find:

$$\left\langle J_s^2 \right\rangle_{exp} = \frac{\kappa_0 k_B T}{c\Phi_0^2}. \qquad (11)$$

This expression for $\left\langle J_s^2 \right\rangle_{exp}$ is valid for frequencies below the vortex depinning frequency, $\omega_p \equiv \kappa_0/\eta$. Vortex motion resulting from fluctuations at higher frequencies is damped by the viscosity, $\eta$, and is negligible. In other words, as a detector of supercurrents, a vortex is a low-pass filter with a bandwidth of $\omega_p/4$.

It is useful to calculate the supercurrent density noise power per unit bandwidth at frequencies below $\omega_p$:

$$S_J(0) = \frac{<J_s^2>}{\omega_p/4} = \frac{4\eta k_B T}{c\Phi_0^2}. \qquad (12)$$

Finally, let us replace $\eta$ with[25] $\Phi_0^2/2\pi\rho_n\xi_{ab}^2$ and calculate the noise power for the sheet current density, $K \equiv cJ_s$:

$$c^2 S_J(0) = \frac{2k_B T}{\pi R_n \xi_{ab}^2}, \qquad (13)$$

where $1/R_n = c/\rho_n$ is the sheet conductance of a unit cell layer. This is the classical result for noise supercurrents in two dimensions, being linearly proportional to the sheet conductance and T.

From the foregoing, we conclude that supercurrent fluctuations have their full classical amplitude for frequencies below $\omega_p$. To determine when quantum effects should become significant, we use the experimental values $\eta = 1\times10^{-6}$ Ns/m$^2$ (which is consistent with the expression for $\eta$ used above) and $\kappa_0 = 2\times10^5$ N/m$^2$ which results in $\omega_p \approx 2\times10^{11}$ rad/s, and $\hbar\omega_p/k_B \approx 1.5$ K. For T > 5 K, our lowest measurement temperature, all of the frequencies in the experimental bandwidth of $\omega_p/4$ are excited at the classical level. At temperatures below 1.5 K, only a fraction, $k_BT/\hbar\omega_p$, of the important bandwidth would be excited, and thus $<J_s^2>_{exp} = S_J(0)k_BT/\hbar\omega_p \propto T^2$. On this basis, we predict that $\kappa_{exp}(T)$ is quadratic rather than linear below about 1.5 K in PLD YBCO films.

## VII. Magnetic Field Dependence of the Labusch Parameter

The magnetic field dependence of the inductivity is complicated by several factors. The field dependent measurements presented here were taken by cooling the film in liquid helium and ramping the field slowly from zero to 6 T. As the field increases, vortices enter the film along paths of least resistance such as grain boundaries or other regions of weak pinning, resulting in a nonuniform vortex density. As the field increases, the vortices are driven deeper into the film and form a lattice with a nearly uniform density. At high fields, the lattice is distorted and intervortex repulsion becomes important. In addition, the film is at all times near the critical state in which the vortices are pushed toward the edges of their pinning sites.

Figure 6 shows $1/\lambda^2_{eff}$ vs. $B_e$ at T = 4.2 K for both films. The discrepancy in $1/\lambda^2$(T = 4.2 K, $B_e$ = 0 T) between Fig. 1a and Fig. 6 occurs because the two measurements were taken 11 months apart and the film changed slightly during that time. There is a drop in $1/\lambda^2_{eff}$ at a few hundred Gauss then a plateau which blends into a gentle decrease in $1/\lambda^2_{eff}$ above about 2 Tesla. As the first vortices enter the film, they only penetrate into the outer edges of the film and have little effect on the measurement. As the field increases above a few hundred Gauss, the vortices penetrate to the center of the film along paths of weak pinning, resulting in the sharp drop in $1/\lambda^2_{eff}$. (Note that this explains why the 1 Tesla curve is out of sequence in Figs. 2b and 4b.) Increasing the field to about 2 Tesla pushes vortices deep into the film where they are pinned by linear defects.

Our analysis focuses on data above 3 Tesla where the vortex density is nearly uniform. Note that the dotted curves in Fig 6, which come from the analysis below, fit the high field data very well and extrapolate to the measured value of $\lambda^{-2}_{eff}$ at zero field. It is upon this observation that we base our assessment that the vortex density is nearly uniform at high fields. Figure 7 shows $\lambda_C^2/B_e$ *vs.* $B_e$ and Fig. 8 shows $\kappa_{exp}$ *vs.* $B_e$.

Through careful consideration of the repulsive forces between vortices in a lattice, it can be shown[26] that the net force felt by a vortex is $F_L = \kappa_L(B)x$ where x is the displacement of the vortex from is lattice position and $\kappa_L(B)$ is an effective Labusch parameter given by $\kappa_L = \dfrac{B_e \Phi_0}{2\mu_0 \lambda^2_{ab}}$. With this result and the relatively minor change in $\kappa_{exp}$ between 2 and 6 Tesla, it can be further shown[26] that each vortex is independently pinned. If any significant portion of the vortices were unpinned, the measured Labusch parameter would be significantly smaller than that observed.

To calculate the field dependence of the Labusch parameter, we assume that a typical vortex is located in a pinning site which lies a distance $x_d$ from the vortex's proper position in the lattice. If each vortex is displaced in a random in direction, the net force felt by each vortex resulting from interactions with all other vortices is $F_L = \kappa_L(B) x_d$. This force pushes the vortex off the center of its pinning site by a distance $\rho_0$ given by:

$$\rho_0 \approx \frac{\kappa_L x_d}{\kappa_0} + O\left(\left(\frac{\kappa_L}{\kappa_0}\right)^3\right) \approx \frac{B_e \Phi_0 x_d}{2\mu_0 \lambda_{ab}^2 \kappa_0}. \qquad (14)$$

At this new location, the effective Labusch parameter is:

$$\kappa_{eff} = \kappa_0 \left[1 - \frac{1}{3}\left(\frac{\rho_0}{\xi_{ab}}\right)^2\right] \approx \kappa_0 - \frac{B_e^2 \Phi_0^2}{12\mu_0^2 \lambda_{ab}^4 \kappa_0} \frac{x_d^2}{\xi_{ab}^2} = \kappa_0\left[1 - \left(B_e/B_0\right)^2\right], \qquad (15)$$

where $B_0 = \frac{2\sqrt{3}\mu_0 \lambda_{ab}^2 \kappa_0 \xi_{ab}}{\Phi_0 x_d}$. $B_0$ is the field at which the force resulting from the strained lattice equals the maximum pinning force. When $B_e$ exceeds $B_0$, the intervortex repulsion will exceed the pinning force and many of the vortices will become unpinned to reduce the strain in the lattice. The vortex inductance should increase dramatically as the external field approaches $B_0$.

The dotted lines in Fig. 8 are fits of Eq. (15) to the data using $\kappa_0 \approx 2 \times 10^5$ N/m$^2$ and $B_0 \approx 15$ T. These values of $\kappa_0$ and $B_0$ fit the data very well for B > 3 Tesla. To test the model more thoroughly requires higher fields. The value $B_0 = 15$ Tesla implies $x_d \approx$ 21 Å, i.e. each vortex is able to find a pinning site within 21 Å of its preferred lattice site.

$\rho_0 \approx 0.7\,\xi_{ab}(0) \approx 10$ Å at 6 Tesla. For comparison, in a triangular lattice at $B_e = 6$ T, the lattice constant is a = 200 Å. These numbers are the *ex posteriori* justification for analyzing the T dependence of the data at 1 Tesla without including the effects of vortex interactions. Clearly these interactions are needed in the theory at higher temperatures even for 1 Tesla.

As stated earlier, the film is near the critical state. Vortices can only enter the film by pushing the vortices already present deeper into the film. As a result, all vortices are pushed toward the edges of their pinning sites and there is a slight gradient in the vortex density. When the film is field cooled, the critical state still exists, but the gradient in the vortex density is in the opposite direction as screening currents attempt to push excess vortices out of the film. Presumably, this is why no difference is observed between field cooled and zero-field cooled measurements.

To determine the effect of the critical state upon our measurement and check our field dependence model at low fields, we have added a small solenoid around the two coil apparatus but within the supperconducting solenoid. By applying a 60 Hz AC current to this solenoid, we can add a small sinusoidal magnetic field to the static magnetic field. The effect of this field is to push vortices deeper into the film and withdraw them. By slowly reducing the AC current to zero, the vortices are expected to be left in a more relaxed state with a much smaller density gradient. Preliminary measurements using this technique extend the range over which our field dependence fits the data to below 50 Gauss. It also shows a 20% increase in the inductivity of the film. This translates into a 50% increase in $\kappa_0(T=0,B=0)$. Thus the critical state has a significant effect on the measured inductivity. More work is being done to better quantify this effect.

VIII. Conclusion

The resistivity of pinned vortices at $T \ll T_g$ provides answers to a number of important questions regarding vortex pinning in YBCO films and the intrinsic properties of the film itself. Their areal density of pinning sites is sufficient to accommodate all of the vortices generated by a field of 6 Tesla. In a model where the linear defects act like cylindrical holes, they have a radius of 3 Å to 8 Å. The vortex pinning potential has a range roughly equal to the ab-plane Ginzburg-Landau coherence length. At its inflection point, the pinning potential has a slope which corresponds to $J_C \approx 10^7$ A/cm$^2$.

We predict a new pinning regime for fields, $B_e > \sim15$ Tesla, where the forces resulting from the distortion of the vortex lattice are sufficient to depin some vortices. This pinning regime should occur before the field is sufficient to fill all available pinning sites. In this regime, the inductivity of the vortices is dominated by the very weak pinning created by the vortex lattice.

The linear temperature dependence of the vortex inductivity at low temperatures shows that classical, thermally induced supercurrent fluctuations dominate the T dependence of $\kappa_{exp}$ from 5 K to $T_g$. The correlation length along the c axis for supercurrents in the ab-plane must be less than the fluctuation length, $\ell_z(0) = 8$ Å to 20 Å at 5 K. We predict that when T drops below 1.5 K, there is a crossover from T to T$^2$ behavior in the Labusch parameter as quantum mechanics freezes out fluctuations below the depinning frequency. Further work is being done to observe this crossover.

For temperatures above 15 K, $\kappa_{exp}(T)$ decreases far slower than can be explained by our model. Some, if not all, of this discrepancy results from the fact that our measurements were made on films near the critical state. If the lattice were fully relaxed, $\kappa_0(T=0,B=0)$ would be larger than observed and $T_g$ of 80 K would be more reasonable. Further work is being done to determine the importance of the critical state in these measurements.

## IX. Acknowledgements

Dr. Rand Biggers at Wright-Patterson AFB kindly provided the PLD films used in this study. We are grateful to James Baumgardner, Brent Boyce, and Stefan Turneaure for providing valuable technical assistance. This work was supported by grants AFOSR 91-0188 and F49620-94-1-0274 and by Dept. Of Energy grant No. DE-FG02-90ER45427 through the Midwest Superconductivity Consortium.

FIGURE CAPTIONS

**Figures 1a & 1b -** Effective penetration depth vs. temperature for films 1 and 2. The top curve is $B_e = 0$ Tesla. The curves then descend from 1 to 6 Tesla in order. For $B_e = 0$, $\lambda_{eff} = \lambda_{ab}$, the magnetic penetration depth of the sample, and $1/\lambda^2$ is proportional to the superfluid density. Data were taken between 5 K and 100 K. The dotted lines below 5 K are extrapolations created using a second order polynomial fit to the data between 5 K and 30 K.

**Figure 2** - Campbell penetration depth vs. temperature. $\lambda_C^2/B_e$ is proportional to the inductance per vortex of the pinned vortices. For film 1, the bottom curve is $B_e = 1$ Tesla with the curves then ascending from 2 to 6 Tesla. Film 2 is identical except the 1 Tesla curve is out of sequence and is the top curve. Data were taken between 5 K and 100 K. The dotted lines below 5 K are extrapolations based on the extrapolations of Fig. 1.

**Figure 3 -** Vortex glass melting curves. Below the curve, the vortices are pinned and their response to the AC field is inductive. Above the curve, the vortices are unpinned and their response to the AC field is dissipative. The solid lines are curve fits showing that $B_g \propto (1-T_g/T_C)^{4/3}$ as observed by other groups. $T_g$ is assigned to be the temperature of the peak in the dissipative component of the mutual inductance.

**Figures 4a & 4b** - Experimental Labusch parameter, $\kappa_{exp}$, vs. temperature. $\kappa_{exp}$ is the measured linear restoring force constant per unit length of the pinned vortices. In Fig. 4a the top curve is $B_e = 1$ Tesla with the curves then descending from 2 to 6 Tesla. Figure 4b is identical except the 1 Tesla curve, the bottom curve, is out of sequence. Data were taken between 5 K and 100 K. The dotted lines below 5 K are extrapolations based on the extrapolations in Fig. 2. The dashed lines are the intrinsic Labusch parameter, $\kappa_0$,

calculated using the measured zero-filed penetration depth and the temperature dependent coherence length calculated by Ulm et al[24].

**Figure 5** – Labusch parameter vs. Temperature. The solid line is $\kappa_{exp}$(T,B=1T), the experimental Labusch parameter taken from Fig. 4a. The dashed line is $\kappa_0$(T), the intrinsic Labusch parameter of the pinning site. $\kappa_0$(T) is calculated using a temperature dependence of $\kappa_0(T) \sim 1/\lambda_{ab}^2(T)\xi_{ab}^4(T)$ and assuming that 1 Tesla is a sufficiently small field that the $\kappa_{exp}$ can be approximated using the isolated vortex model. The dotted line is $\kappa_{eff}$(T) as predicted by our model.

**Figure 6** – Effective penetration depth vs. magnetic field. $1/\lambda^2_{eff}$ is proportional to the imaginary resistivity of the film. The dotted line is the effective penetration depth predicted by our model.

**Figure 7** – Campbell penetration depth vs. magnetic field. $\lambda_C^2/B_e$ is proportional to the inductance per vortex. The dotted line is the Campbell penetration depth predicted by our model.

**Figures 8** – Labusch parameter, $\kappa_{exp}$, vs. magnetic field. The Labusch parameter is the linear restoring force constant for the pinned vortices. The dotted curve is the Labusch parameter predicted by our model.

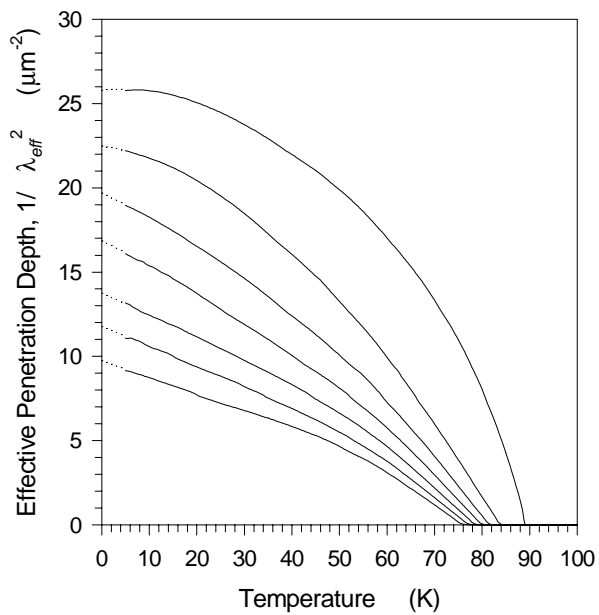
Figure 1a

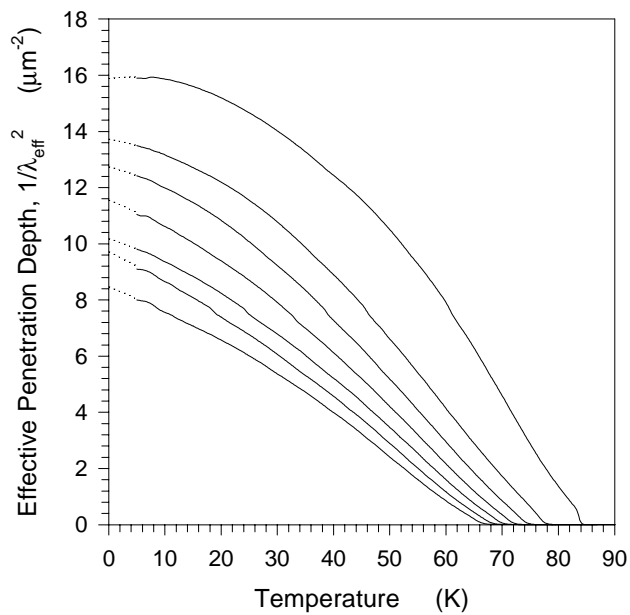
Figure 1b

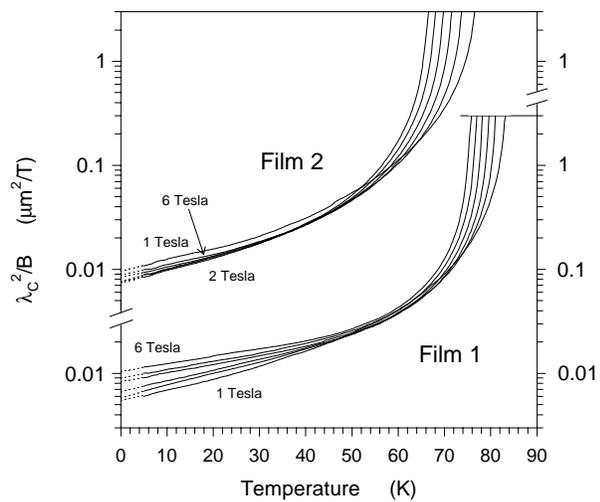
Figure 2

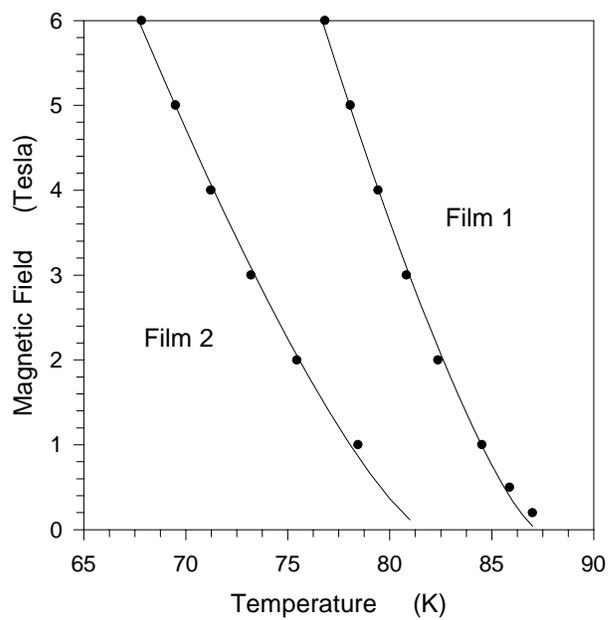
Figure 3

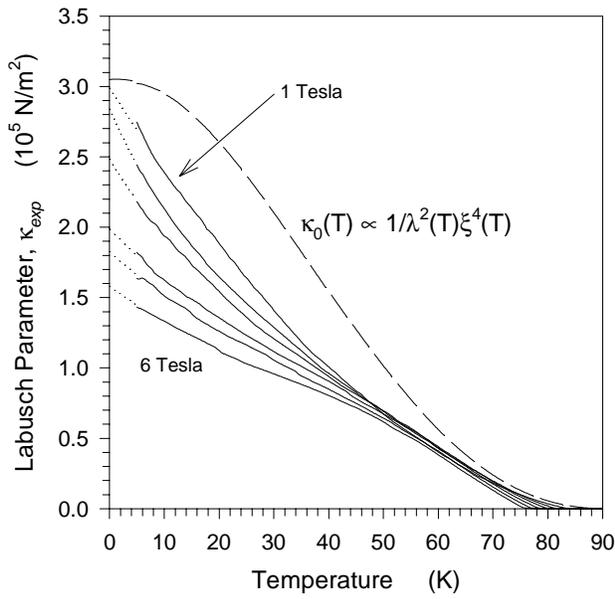
Figure 4a

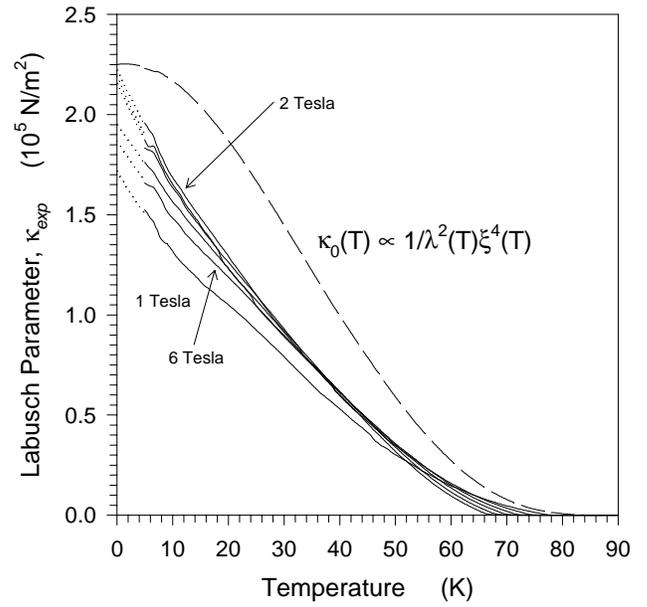
Figure 4b

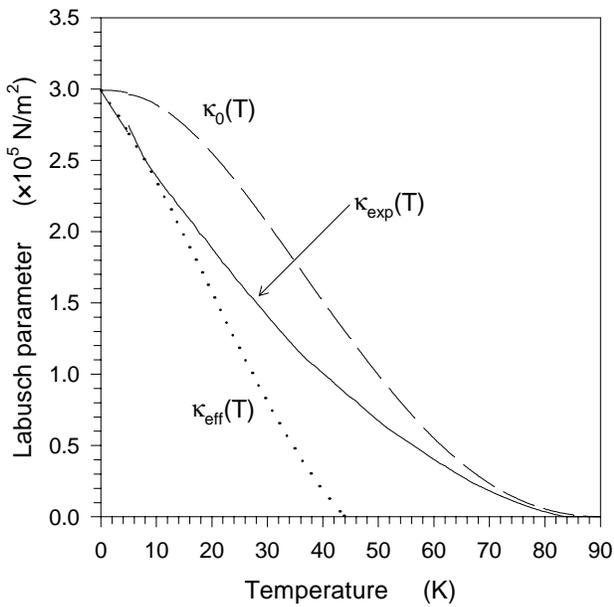
Figure 5

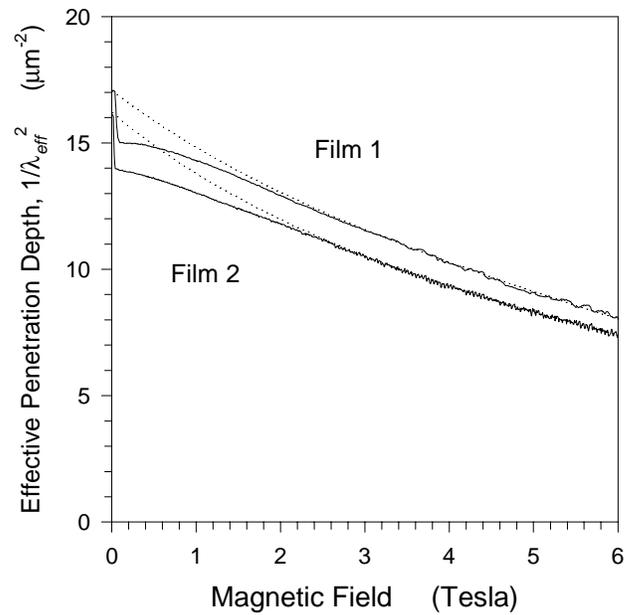
Figure 6

Figure 7

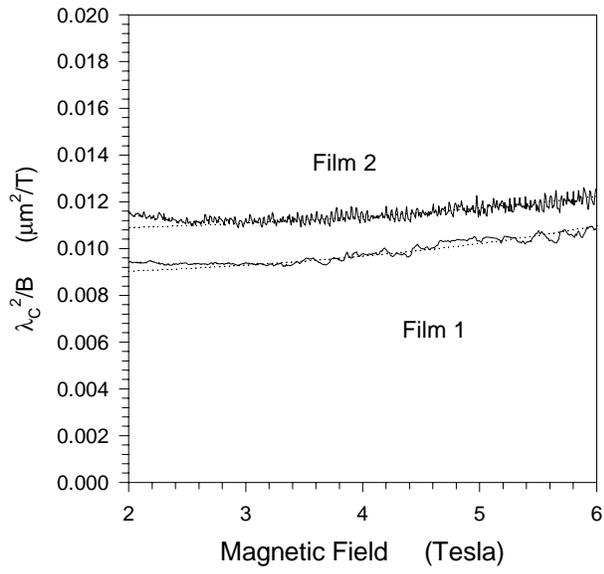

Figure 8

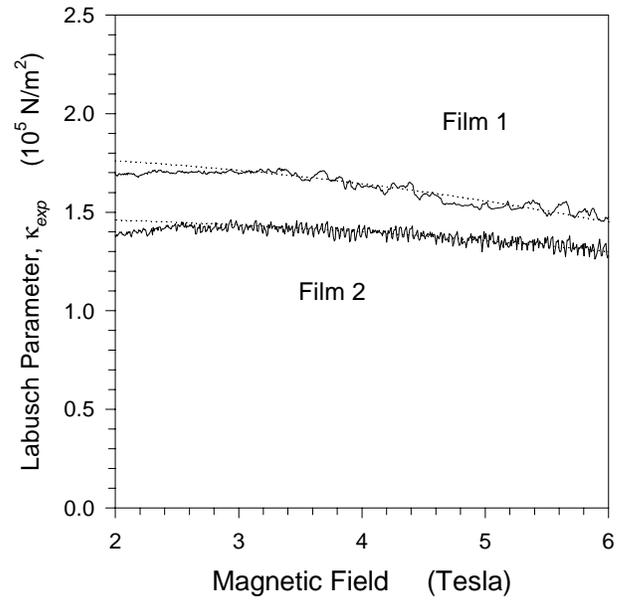